# A Robust Method for Pitch Tracking in the Frequency Following Response using Harmonic Amplitude Summation Filterbank

**Sajad Sadeghkhani[1]**, Maryam Karimi Boroujeni[2], Hilmi R. Dajani[1], Saeid R. Seydnejad[1,3], Christian Giguère[2]

1. School of Electrical Engineering and Computer Science, Faculty of Engineering, University of Ottawa, Ottawa, Canada.
2. School of Rehabilitation Sciences, Faculty of Health Sciences, University of Ottawa, Ottawa, Canada.
3. Department of Electrical Engineering, Shahid Bahonar University of Kerman, Kerman, Iran.

**Abstract**

The Frequency Following Response (FFR) reflects the brain's neural encoding of auditory stimuli including speech. Because the fundamental frequency (F0), a physical correlate of pitch, is one of the essential features of speech, there has been particular interest in characterizing the FFR at F0, especially when F0 varies over time. The standard method for extracting F0 in FFRs has been the Autocorrelation Function (ACF). This paper investigates harmonic-structure-based F0 estimation algorithms, originally developed for speech and music, and resolves their poor performance when applied to FFRs in two steps. Firstly, given that unlike in speech or music, stimulus F0 of FFRs is already known, we introduce a stimulus-aware filterbank that selectively aggregates amplitudes at F0 and its harmonics while suppressing noise at non-harmonic frequencies. This method, called Harmonic Amplitude Summation (HAS), evaluates F0 candidates only within a range centered around the stimulus F0. Secondly, unlike other pitch tracking methods that select the highest peak, our method chooses the most prominent one, as it better reflects the underlying periodicity of FFRs. To the best of our knowledge, this is the first study to propose an F0 estimation algorithm for FFRs that relies on harmonic structure. Analyzing recorded FFRs from 16 normal hearing subjects to 4 natural speech stimuli with a wide F0 variation from 89 Hz to 452 Hz showed that this method outperformed ACF by reducing the average Root-Mean-Square-Error (RMSE) within each response and stimulus F0 contour pair by 8.8% to 47.4%, depending on the stimulus.

## 1. Introduction

Pitch is an important auditory feature that plays a crucial role in how we interpret and experience sound. Despite having a perceptual definition, it is often described in terms of its physical correlates such as the signal periodicity or its inverse the fundamental frequency F0. In this study, we do not differentiate between fundamental frequency and pitch. Pitch is present in a wide variety of acoustic signals, including speech and music, making pitch extraction an important focus in numerous fields. Applications range from speech and music analysis (De Medeiros et al., 2021) to medical diagnosis (Li et al., 2023), and linguistic research (Singh & Fu, 2016).

The Frequency Following Response (FFR) is a neurophysiological scalp-recorded response that reflects how the central auditory system encodes acoustic stimuli. One of the key features studied in these responses is the component at F0. While a number of studies have employed synthetic speech stimuli, there is an increasing interest in employing natural speech stimuli in which F0 varies over time (Bachmann et al., 2021; Dajani et al., 2005; Krishnan et al., 2004), highlighting the increasing importance of accurate pitch tracking. How well the F0 in the FFR follows the stimulus F0 for different speakers and listeners, including those with hearing impairment, is still unknown and could be an important diagnostic marker of auditory processing.

FFRs have very small amplitudes and are often obscured by background noise. To mitigate this, response sweeps are repetitively recorded and coherently averaged, a process that often requires long recording sessions. Xu and Ye (2014) introduced a posteriori Wiener filtering, which enhanced FFR extraction and outperformed the conventional averaging method; however, it did not directly focus on pitch tracking within FFRs. Recently, methods have been developed to assess the strength of the response at F0 in continuous natural speech (Forte et al., 2017; Maddox & Lee, 2018). However, these methods do not track the evolution of the response at F0 over time. Developing advanced pitch tracking techniques specifically tailored for FFR analysis could significantly reduce recording times, making the process more feasible for all clinical populations, particularly infants who struggle to remain still during extended sessions.

Pitch tracking in speech has been the subject of extensive research. Many algorithms designed for this task can be classified into parametric and non-parametric methods. Non-parametric methods can be further divided into time-domain and frequency domain approaches. As a time-domain approach, the Autocorrelation Function (ACF) has long been used to detect pitch in speech (Rabiner, 1977). ACF F0 contours were smoothed using Viterbi algorithm and incorporated into Praat (Boersma & Van Heuven, 2001), enhancing their robustness in noise. In the frequency domain, notable algorithms include SWIPE (Camacho & Harris, 2008) and PEFAC (Gonzalez & Brookes, 2014). Both methods rely on the harmonic structure of voiced speech, using the summation of harmonic energy as the core principle of their F0 estimation. As for parametric methods, a state-of-the-art noise-resilient algorithm follows a Bayesian estimation approach (Shi et al., 2019). Recently, Wei et al. (2022) implemented the idea of logarithmic harmonic summation in a multiple-rates dilated convolution neural network to estimate speech F0.

Despite significant advancements in robust pitch tracking methods for speech, limited research has focused on developing similar techniques for the Frequency Following Response (FFR), a neurophysiological scalp-recorded response that reflects how the central auditory system encodes audio stimuli. Most papers that explore the F0 contour in the FFR apply ACF to extract the dominant periodicity of the response (Krizman & Kraus, 2019). This observation is not surprising: while state-of-the-art speech pitch tracking methods outperform ACF in noisy speech, they fall short when applied to FFRs. We believe there are three underlying reasons for this. First, ACF benefits from prior knowledge of the stimulus F0 and is typically applied only to pitch periods near this frequency(Jeng et al., 2011), whereas speech pitch tracking methods search across a wide frequency range without incorporating such prior information. Second, in FFRs, most of the signal power—including both harmonic energy and noise—is concentrated at low frequencies. In contrast, speech signals often contain stronger harmonic content near formants, which are not necessarily low in frequency. As a result, spectral or harmonic normalization techniques commonly used in speech F0 estimation may not be effective for FFRs. Third, the harmonic structure in FFRs is weaker than in speech signals. While 8 to 10 harmonics are often analyzed in speech, FFRs typically contain far fewer. Including additional harmonics in the analysis of FFR tends to introduce more noise than usable signal energy, ultimately degrading performance.

We mitigate these three limitations by proposing a filterbank designed to aggregate the amplitude of the fundamental frequency and its first few (2 to 4) harmonics. Unlike Gonzalez and Brookes (2014) and Wei et al. (2022) who employ a logarithmic-scale spectrogram, our method utilizes a Discrete Fourier Transform (DFT) on a linear frequency scale. One of the main reasons that log-scale is more effective in speech pitch tracking is that it makes the distance between harmonics independent of the F0 value reducing the chance of octave errors in F0 estimation. However, this reduces frequency resolution especially at high frequencies. In contrast, because the F0 of the stimulus in the FFR is known, and assuming the response F0 does not significantly deviate from the stimulus F0 (e.g., remain within an octave), we can constrain the F0 search range, ensuring that other harmonic peaks in the filtered response spectrum do not interfere with F0 estimation while preserving frequency resolution. Finally, the peak-picking algorithm in our method selects the most prominent peak rather than simply the highest one. This prominence-based approach is less influenced by the overall downward slope of the spectrum, which is largely shaped by the noise power distribution.

## 2. Method

*2.1 Participants*

This study utilizes data collected by Karimi Boroujeni et al. (2024). Participants included 16 adults (8 males, 8 females) aged 18-31 who all underwent hearing tests to confirm normal hearing sensitivity. Participants showed normal hearing across key frequencies, with thresholds under 25 dB HL bilaterally from 250 to 8000 Hz. All procedures were performed in compliance with relevant laws and guidelines of the University of Ottawa and have been approved by the institutional research ethics board (Reference number: H-03-22-7915, July 22, 2022). The privacy rights of the participants have been observed, and their informed consent was obtained.

*2.2 Stimuli*

The stimuli used to evoke FFRs consisted of the word "balloon", articulated by a male and female speaker with sad and happy emotions. This word was derived from the sentence "No, I burst the balloon!" in the Emotional Speech Database (ESD)(Zhou et al., 2022), and the stimuli were labeled as Male Sad (MS), Male Happy (MH), Female Sad (FS), and Female Happy (FH). The recordings were sampled at 44,100 Hz, with durations ranging from 460 ms to 660 ms (Karimi Boroujeni et al., 2024).

To extract the F0 contours of the stimuli, we employed the built-in 'pitch' and 'pitchnn' commands in MATLAB R2024a, which offer multiple pitch estimation methods. Each stimulus was first segmented into 50 ms frames with a 10 ms step size, and the F0 contour was estimated for each frame using all available methods. During the extraction, small discrepancies of a few hertz were observed between methods in some frames, along with occasional octave errors. To mitigate these issues, the median F0 value across all methods was chosen as the final F0 estimate for each frame. A few octave errors were manually corrected in the final step.

The stimuli were selected to contain a wide range of F0 values for pitch tracking analysis. The F0 contour for MS displays a nearly continuous falling pattern, starting at 111 Hz and descending to 85 Hz. For MH, the F0 contour begins at 110 Hz, rises to 181 Hz, and then decreases to 164 Hz. The FS F0 curve starts at 280 Hz, rises to 331 Hz, and then falls to approximately 191 Hz. Similarly, the FH F0 curve begins at 304 Hz, peaks at 437 Hz, and then drops to 180 Hz. Stimuli were delivered binaurally via shielded insert earphones at 80 dB SPL and a presentation rate of 1.26 stimuli/sec, with alternating polarity. The order of stimuli presentation was randomized across participants.

*2.3 Responses*

The FFR was recorded using the Duet 2-channel Auditory Evoked Potential (AEP) system (Intelligent Hearing Systems, Miami, FL), with four disposable Ag/AgCl disc electrodes. Inverting electrodes were placed on the left (A1) and right (A2) earlobes following the International Electrode System (IES) 10-20 guidelines, the noninverting electrode was positioned at the vertex (Cz), and the ground electrode was placed on the forehead (Fz). To ensure optimal contact, the electrode sites were prepped using alcohol wipes and Nuprep Skin Prep Gel. The electrode impedances threshold was set to 5 kΩ, with inter-electrode impedance differences not exceeding 3 kΩ (Karimi Boroujeni et al., 2024)

Participants were seated comfortably in a soundproof and electrically shielded booth during the recordings. They were instructed to remain still and close their eyes. Due to the length of the recordings, the experiment was conducted in two sessions, each lasting 3 hours. During each session, 1500 sweeps (three blocks of 500) were recorded for each condition, resulting in 3000 sweeps per condition across both sessions. The FFRs were recorded with a 175 $\mu s$ sampling interval and were bandpass filtered online between 30 and 1500 Hz (Karimi Boroujeni et al., 2024).

*2.4 F0 Tracking Algorithm*

We frame the response to 50 ms segments with 10 ms overlap. Each frame is considered a periodic signal $x(n)$ with harmonic structure plus additive noise $e(n)$ as follows

$$x(n) = \sum_{k=1}^{K} \alpha_k \cos(j2\pi k F0 n + \varphi_k) + e(n), \quad 0 \leq n \leq L \quad (1)$$

$$x = [x(1)\ x(2)\ x(3)\ ...\ x(L)] \quad (2)$$

where $\alpha_k$ and $\varphi_k$ are the magnitude and phase of the $k^{th}$ harmonic, respectively, with $K$ total harmonics. Frame $x$ consists of $L$ samples. Each filter $h_{F0_i}(n)$ in every branch of the filterbank can be similarly defined as,

$$h_{F0_i}(n) = \sum_{k=1}^{K} \cos(j2\pi k F0_i n), \quad F0_{min} \leq F0_i \leq F0_{max} \tag{3}$$

$$\boldsymbol{h}_{F0_i} = [h_{F0_i}(1)\ h_{F0_i}(2)\ h_{F0_i}(3)\ ...\ h_{F0_i}(L)] \tag{4}$$

where $i$ denotes different $F0$ candidates in the range of $F0_{min} = 80Hz$ and $F0_{max} = 500Hz$. For simplicity, each complex tone in Eq. (3) has a unity magnitude and zero phase.

By taking the Discrete Fourier Transform (DFT) of $\boldsymbol{h}_{F0_i}$, we can construct the matrix $\mathbf{H}$ as

$$\mathbf{H} = [\boldsymbol{F}_{h_1}\ \boldsymbol{F}_{h_2}\ \boldsymbol{F}_{h_3}\ ...\ \boldsymbol{F}_{h_M}]^T \tag{5}$$

$$\boldsymbol{F}_{h_i} = |\mathcal{F}\{\boldsymbol{h}_{F0_i}\}^T| \tag{6}$$

where $\mathcal{F}\{\cdot\}$ denotes the DFT operation and $\boldsymbol{F}_{h_i} \in \mathbb{R}^{1 \times N}$ and $\mathbf{H} \in \mathbb{R}^{M \times N}$. $M$ denotes the number of filters selected at 1 Hz increments (i.e. $M = F0_{max} - F0_{min} + 1$) and $N$ represents the total number of frequency components. We padded the filters in Eq. (4) and response frames in Eq. (2) with $(2N - L)$ zeros to obtain $N$ non-negative frequency components. Applying the filterbank defined in Eq. (5) to $|\mathcal{F}\{\boldsymbol{x}\}|$ produces M filtered frames. Among these, one output frame contains the harmonic amplitudes, while the remaining filtered frames primarily consist of noise or partial harmonics. (e.g. if F0=100 Hz, $F0_{71}$=150 Hz and $\boldsymbol{h}_{F0_{71}}$ sums amplitudes of the third harmonic at 300 Hz and the sixth harmonic at 600 Hz as well as non-harmonic components at 150 Hz, 450 Hz). We assume that the summation of harmonic amplitudes exceeds the level of the noise. By adding the frequency components within each filtered frame, we obtain

$$\boldsymbol{y} = ([\mathbf{1}_M^T \cdot |\mathcal{F}\{\boldsymbol{x}\}|] \odot \mathbf{H})\mathbf{1}_N^T, \quad \boldsymbol{y} \in \mathbb{R}^{M \times 1} \tag{7}$$

$$\mathbf{1}_M^T = [1\ 1\ 1\ ...\ 1]^T \tag{8}$$

Where $\mathbf{1}_M^T$ is a column vector consisting of $M$ ones and $\odot$ denotes Hadamard (element-wise) product.

The objective is to identify $F0_i$ at which $\boldsymbol{y}$ in Eq. (7) produces the most prominent peak (Kirmse & de Ferranti, 2017) within a defined threshold (50 Hz) around the stimulus F0 of the same time-aligned frame. To do that, we first calculate the difference of $\boldsymbol{y}$ by shifting one sample to the left and subtracting the original $\boldsymbol{y}$ from the shifted vector. Peaks are detected where this difference results in a sign change. Finally, prominence peak-picking performed frame by frame results in the response F0 contour.

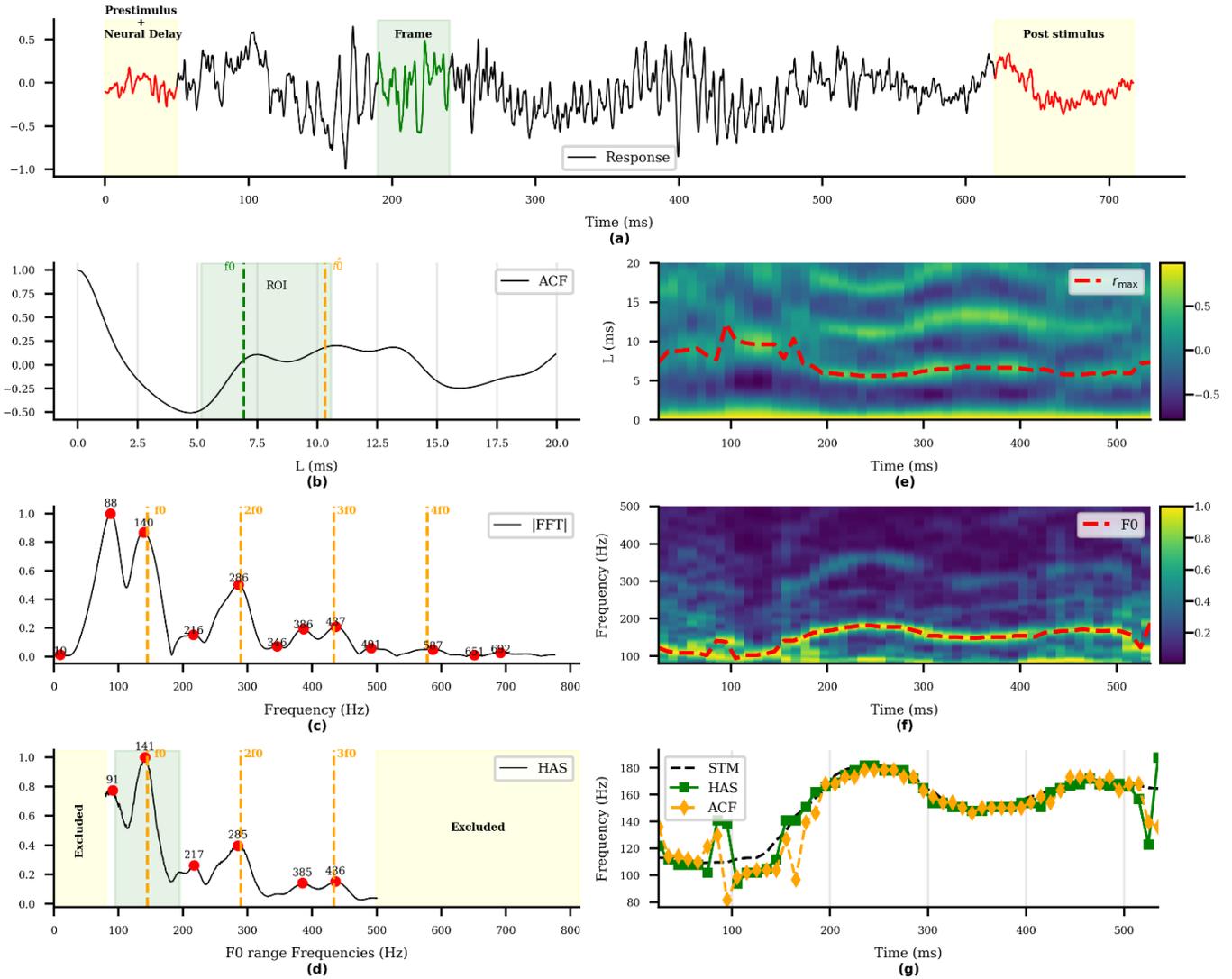

Fig. 1. Comparison of HAS-PR and ACF F0 tracking. (a) FFR of a participant in response to a Male Happy stimulus. (b–d) ACF, absolute DFT, and HAS-PR output of the FFR calculated in Eq. (7) for the highlighted frame in (a), with the F0 search region marked in green. The vertical green line, orange line, and red dots represent the stimulus F0, the estimated response F0, and the peaks detected by the peak-picking algorithm, respectively. (e, f) Correlogram and HAS-PR output across all frames. The red dashed line represents the lag corresponding to the maximum correlation (estimated pitch period) and the estimated F0 contours across all frames, respectively. (g) Estimated F0 contours from the ACF and HAS-PR methods, along with the stimulus F0 contour.

The F0 tracking procedure using Harmonic Amplitude Summation with prominence peak-picking (HAS-PR) and its comparison with ACF is displayed in Fig. 1. In Fig. 1(a), the initial 50 ms of the response corresponding to 40 ms of pre-stimulus activity and a 10 ms neural delay, is removed. The remaining response is segmented into frames using the same window length and step size as the stimulus to ensure precise time alignment. Framing continues until the number of response frames matches that of the stimulus frames, after which framing stops, and post-stimulus recording is discarded. While ACF fails to capture the pitch period in Fig. 1(b), the absolute DFT in Fig. 1(c) shows a peak at F0 (140 Hz), although it cannot be easily distinguished from its adjacent higher peak at 88 Hz. The HAS-PR output in Fig. 1(d), however, contains a peak very close to F0 that stands out prominently due to the presence of strong harmonic components.

## 3. Results

### 3.1 Comparison of HAS performance with other methods

Fig. 2 compares the FFR F0 contours from three participants obtained by HAS-PR and ACF. This figure highlights the inaccuracies of the ACF in tracking pitch under noisy conditions. Moreover, it is observed that both methods more accurately track male-uttered stimuli (stimuli with lower mean F0 value) and the differences between the two methods become more evident for female talkers with a higher mean F0. Notably, HAS-PR performs better in all conditions.

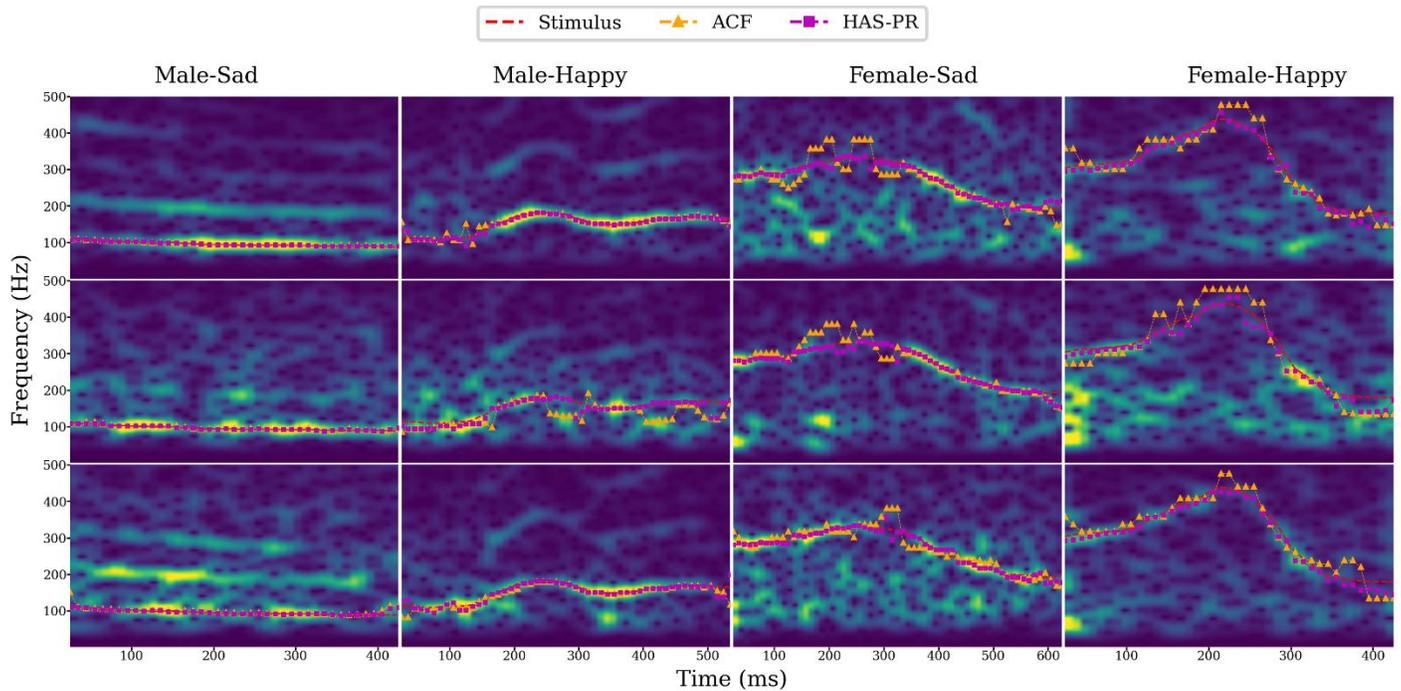

Fig. 2. Example FFR F0 contours estimated by HAS-PR and ACF along with stimulus F0 contours for three participants and four stimulus emotion conditions. The background shows the spectrogram of the averaged response.

The performance of HAS was further compared with several established F0 estimation algorithms in addition to ACF, including Praat (Jadoul et al., 2018), PEFAC (Gonzalez & Brookes, 2014) without amplitude compression, HarmoF0 (Wei et al., 2022), and Bayesian pitch tracker (Shi et al., 2019). Fig. 3 displays RMSE values between stimulus and estimated response F0 contours for 16 subjects, plotted against the number of averaged sweeps. In this figure, "HAS-PR" denotes the HAS method using prominence peak picking, while "HAS-HT" indicates that peaks with maximum height were selected.

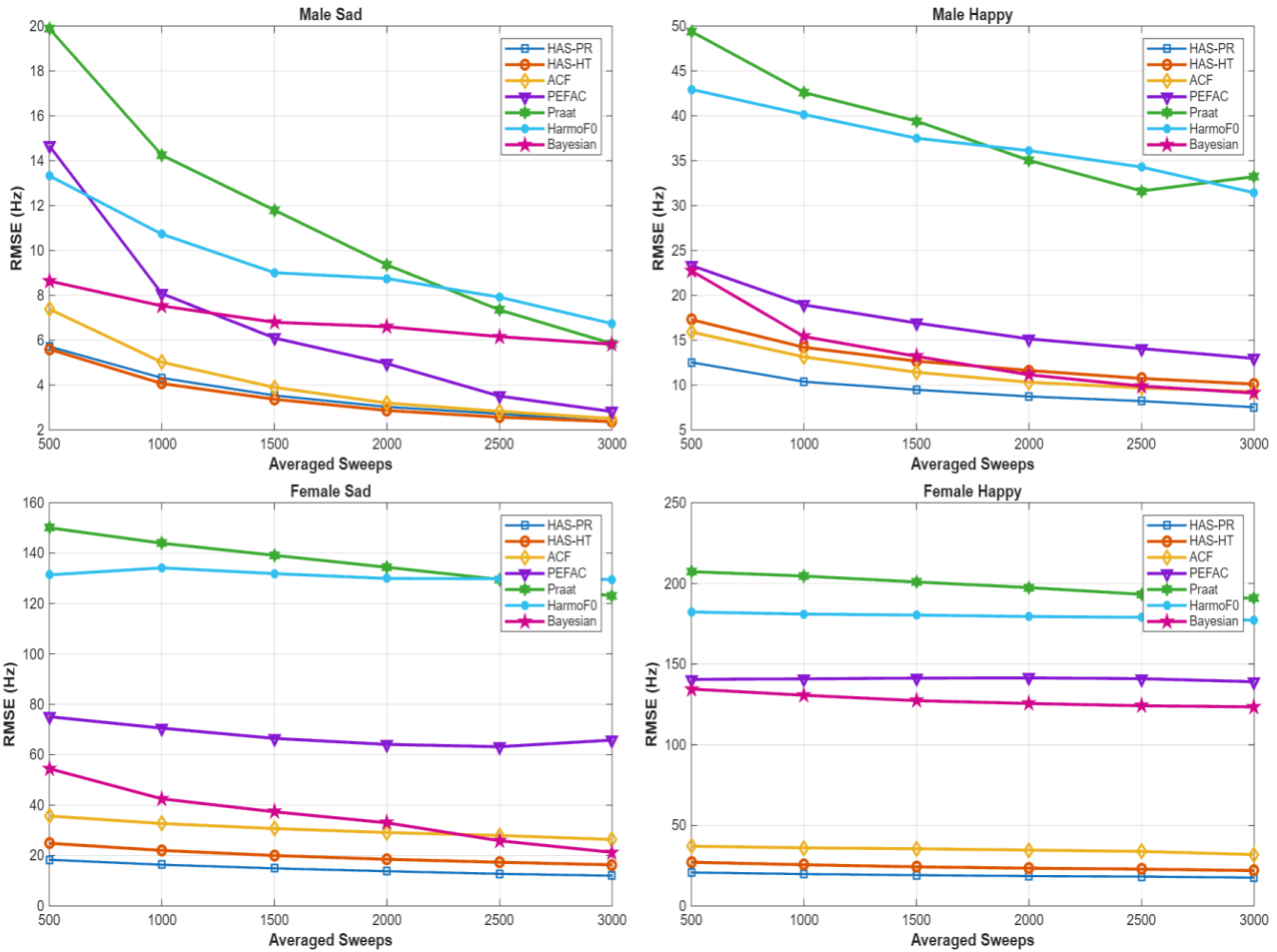

Fig. 3. RMSE between stimulus and response F0 contours versus the number of averaged sweeps for the four stimulus conditions (Male Sad, Male Happy, Female Sad, and Female Happy).

To obtain the F0 contours, the response was band-pass filtered, with the lower cutoff set to 20 Hz below the minimum F0 frequency of the corresponding stimulus and the upper cutoff set to 2000 Hz. This cutoff was chosen because phase-locking in auditory neurons, the neurophysiological basis of the FFR, is likely too weak to produce measurable components in the response at higher frequencies. Because the speech-based methods included a voiced/unvoiced decision block, parameters were adjusted so that all frames were treated as voiced. This ensured that the algorithms consistently provided their best F0 estimate, instead of failing to return an F0 value for frames that would otherwise be classified as unvoiced.

Table 1 presents the mean and standard deviation (SD) of RMSEs between stimulus and response F0 contours across participants. It also reports the Gross Pitch Error (GPE), defined as the number of frames where the estimated response F0 deviates by more than 20% from the stimulus F0 (Drugman & Alwan, 2019), and $RMSE_{20}$, which denotes the RMSE for frames without gross pitch errors. We introduced this metric because large pitch errors, such as octave errors, easily inflate the standard RMSE, limiting its ability to fully reflect F0 tracking accuracy. For example, in Table 1, the Bayesian method on the Female Sad shows that only 7.2% of frames contribute to gross pitch errors. These erroneous frames increase the RMSE from 12.6 Hz (frames without gross pitch errors) to 21.2 Hz (all frames).

The findings demonstrate that the HAS-PR outperforms other methods in all stimuli and shows lower RMSE than HAS-HT in 3 out of 4 stimuli, but not in the "Male Sad" stimulus. Both HAS-PR and HAS-HT, however, performed the same for this stimulus when all 3000 response sweeps were averaged. Moreover, the performance difference is particularly pronounced in stimuli with higher F0 values, making HAS-PR very useful for female talker stimuli, in which the harmonic structure in responses was notably weaker than that observed for male talker stimuli. All speech-based methods were susceptible to octave errors, mainly caused by a component at F0/2 or strong noise components at lower frequencies. The only exceptions were the Bayesian method and PEFAC, both of which showed promising performance with the Female Sad stimulus. However, none of these established methods performed well on Female Sad.

Table 1 Comparison of methods based on RMSE (Hz), $RMSE_{20}$ (Hz), and Gross Pitch Error (GPE) between stimulus and response F0 contours for the four stimulus conditions. Means (across all participants) are shown.

| Method | MS | | | MH | | | FS | | | FH | | |
|---|---|---|---|---|---|---|---|---|---|---|---|---|
| | RMSE | $RMSE_{20}$ | GPE | RMSE | $RMSE_{20}$ | GPE | RMSE | $RMSE_{20}$ | GPE | RMSE | $RMSE_{20}$ | GPE |
| HarmoF0 | 6.7 | 3.9 | 3.1 | 31.4 | 8.5 | 21 | 130 | 17.1 | 76 | 177 | 19.5 | 81 |
| Praat | 5.9 | 2.8 | 1.2 | 41 | 7.4 | 24.5 | 123 | 15.7 | 44 | 19 | 18.7 | 73 |
| Bayesian | 3.2 | 2.9 | 0.5 | 9.1 | 6.3 | 4.1 | 21.2 | 12.6 | 7.2 | 123 | 21.9 | 59.4 |
| PEFAC | 2.8 | 1.7 | 0.3 | 13 | 6.3 | 6.9 | 66 | 13.9 | 27 | 144 | 23.8 | 64 |
| ACF | 2.6 | 2.17 | 0.5 | 10.9 | 7 | 6.1 | 19.3 | 18.7 | 2.7 | 33.4 | 31.2 | 11 |
| HAS-HT | **2.37** | **2.15** | **0.16** | 10.1 | **6** | 5 | 16.3 | 13.6 | 3.4 | 21.9 | 19.9 | 5.4 |
| HAS-PR | **2.37** | **2.15** | **0.16** | 7.5 | 6.1 | **1.4** | 12 | **10.7** | 1.2 | 17.57 | 16 | 3 |

## 3.2 Statistical Evaluation of the method Performance

The number of harmonics was analyzed to evaluate HAS-PR performance across different stimulus conditions. The optimal K value in Eq. (3) which resulted in the minimum RMSE in Table 1 was higher for the MS condition, which had a lower F0 contour. This observation indicates that higher stimulus F0 contours lead to weaker response harmonic structures. Specifically, the optimal $K$ for the MS condition was 4, while it was 2 for the rest. To ensure statistical reliability, paired t-tests were performed on three sets of comparisons. First, comparing HAS-PR with optimal $K$ to that with $K = 1$ yielded statistically significant differences (p-values < 0.05) for all conditions except FS (p-value = 0.078). This shows that F0 estimation using harmonic amplitude summation over F0, and one or more harmonics is more accurate than relying on only the component at F0. Second, HAS-PR with optimal $K$ was compared to ACF, resulting in p-values < 0.05 for all conditions except MS. Finally, comparisons between the HAS-PR and HAS-HT methods yielded p-values < 0.05 across all conditions except MS where all three methods produced accurate and similar F0 contours with lower errors compared to the other conditions.

## 4. Discussion

We developed a stimulus-aware pitch tracking method for FFR recordings using a Harmonic Amplitude Summation (HAS) filterbank with prominence peak-picking referred to as HAS-PR. We showed that, compared to ACF and a few well-known pitch tracking methods, this approach enhances F0 estimation by emphasizing harmonic structures and utilizing prominence peak picking.

Since we observed that FFR power exhibits a downward trend with frequency (e.g. Fig. 1(c)), we considered using a detrending step crucial before selecting the highest peak. However, instead of dividing this task into two steps, we proposed prominence-based peak-picking, which is inherently independent of the curve trend. The results presented in Table 1 demonstrate that such a peak better reflects the F0 value.

In conclusion, we introduced HAS-PR, a frequency-domain pitch tracking method in FFRs designed to isolate and aggregate frequencies corresponding to F0 and its harmonics with a focus on frequencies within a neighbourhood around the stimulus F0 value to prevent octave errors. This contrasts with the conventional ACF, which operates in the time domain and lacks an inherent noise suppression mechanism. We evaluated the RMSE and GPE between the F0 contour of the stimulus and those obtained using HAS, ACF, and other methods on 64 FFRs recorded from 16 normal-hearing participants in response to four stimuli. Compared to the ACF, the improvement in mean RMSE was 8.8% for Male Sad, 31.2% for Male Happy, 37.8% for Female Sad, and 47.4% for Female

Happy stimuli. In terms of GPE, improvements of 0.34% in Male Sad, 4.7% in Male Happy, 1.5% in Female Sad, and 8% in Female Happy were observed.

## Acknowledgments

This research was supported by a Discovery Grant (RGPIN-2020-03990) from the Natural Sciences and Engineering Research Council of Canada. Portions of the research in this paper used the ESD Database made available by the HLT lab, National University of Singapore, Singapore.## References

Bachmann, F. L., MacDonald, E. N., & Hjortkjær, J. (2021). Neural measures of pitch processing in EEG responses to running speech. *Frontiers in Neuroscience*, 15, 738408, https://doi.org/10.3389/fnins.2021.738408

Boersma, P., & Van Heuven, V. (2001). Speak and unSpeak with PRAAT. *Glot International*, 5(9/10), 341-347,

Camacho, A., & Harris, J. G. (2008). A sawtooth waveform inspired pitch estimator for speech and music. *The Journal of the Acoustical Society of America*, 124(3), 1638-1652, https://doi.org/10.1121/1.2951592

Dajani, H. R., Purcell, D., Wong, W., Kunov, H., & Picton, T. W. (2005). Recording human evoked potentials that follow the pitch contour of a natural vowel. *IEEE Transactions on Biomedical Engineering*, 52(9), 1614-1618, https://doi.org/10.1109/TBME.2005.851499

De Medeiros, B. R., Cabral, J. P., Meireles, A. R., & Baceti, A. A. (2021). A comparative study of fundamental frequency stability between speech and singing. *Speech Communication*, 128, 15-23, https://doi.org/10.1016/j.specom.2021.02.003

Drugman, T., & Alwan, A. (2019). Joint robust voicing detection and pitch estimation based on residual harmonics. *arXiv preprint arXiv:2001.00459*, https://doi.org/10.48550/arXiv.2001.00459

Forte, A. E., Etard, O., & Reichenbach, T. (2017). The human auditory brainstem response to running speech reveals a subcortical mechanism for selective attention. *Elife*, 6, e27203, https://doi.org/10.7554/eLife.27203

Gonzalez, S., & Brookes, M. (2014). PEFAC-A pitch estimation algorithm robust to high levels of noise. *IEEE/ACM Transactions on Audio, Speech, and Language Processing*, 22(2), 518-530, https://doi.org/10.1109/TASLP.2013.2295918

Jadoul, Y., Thompson, B., & De Boer, B. (2018). Introducing parselmouth: A python interface to praat. *Journal of Phonetics*, 71, 1-15,

Jeng, F.-C., Hu, J., Dickman, B., Montgomery-Reagan, K., Tong, M., Wu, G., & Lin, C.-D. (2011). Cross-linguistic comparison of frequency-following responses to voice pitch in American and Chinese neonates and adults. *Ear and hearing*, 32(6), 699-707, https://doi.org/10.1097/aud.0b013e31821cc0df

Karimi Boroujeni, M., Sadeghkhani, S., Seyednejad, S. R., Dajani, H., & Giguère, C. (2024). The neural representation of emotional cues investigated using the speech frequency following response: A potential tool to evaluate speech prosody. *The Journal of the Acoustical Society of America*, 156(4_Supplement), A53-A53, https://doi.org/10.1121/10.0035084

Kirmse, A., & de Ferranti, J. (2017). Calculating the prominence and isolation of every mountain in the world. *Progress in Physical Geography*, 41(6), 788-802,

Krishnan, A., Xu, Y., Gandour, J. T., & Cariani, P. A. (2004). Human frequency-following response: representation of pitch contours in Chinese tones. *Hearing Research*, 189(1-2), 1-12, https://doi.org/10.1016/s0378-5955(03)00402-7

Krizman, J., & Kraus, N. (2019). Analyzing the FFR: A tutorial for decoding the richness of auditory function. *Hearing Research*, 382, 107779, https://doi.org/10.1016/j.heares.2019.107779

Li, Q., Millard, K., Tetnowski, J., Narayana, S., & Cannito, M. (2023). Acoustic analysis of intonation in persons with Parkinson's disease receiving transcranial magnetic stimulation and intensive voice treatment. *Journal of Voice*, 37(2), 203-214, https://doi.org/10.1016/j.jvoice.2020.12.019

Maddox, R. K., & Lee, A. K. (2018). Auditory brainstem responses to continuous natural speech in human listeners. *Eneuro*, 5(1), https://doi.org/10.1523/eneuro.0441-17.2018

Rabiner, L. (1977). On the use of autocorrelation analysis for pitch detection. *IEEE Transactions on Acoustics, Speech, and Signal Processing*, 25(1), 24-33, https://doi.org/10.1109/TASSP.1977.1162905

Shi, L., Nielsen, J. K., Jensen, J. R., Little, M. A., & Christensen, M. G. (2019). Robust Bayesian pitch tracking based on the harmonic model. *IEEE/ACM Transactions on Audio, Speech, and Language Processing*, 27(11), 1737-1751, 10.1109/TASLP.2019.2930917

Singh, L., & Fu, C. S. (2016). A new view of language development: the acquisition of lexical tone. *Child Development*, 87(3), 834-854, https://doi.org/10.1111/cdev.12512

Wei, W., Li, P., Yu, Y., & Li, W. (2022). Harmof0: Logarithmic scale dilated convolution for pitch estimation. *2022 IEEE International Conference on Multimedia and Expo (ICME)*, https://doi.org/10.1109/ICME52920.2022.9858935

Xu, Q., & Ye, D. (2014). Evaluation of a posteriori Wiener filtering applied to frequency-following response extraction in the auditory brainstem. *Biomedical Signal Processing and Control*, 14, 206-216, https://doi.org/10.1016/j.bspc.2014.08.003

Zhou, K., Sisman, B., Liu, R., & Li, H. (2022). Emotional voice conversion: Theory, databases and esd. *Speech Communication*, 137, 1-18, https://doi.org/10.1016/j.specom.2021.11.006